\documentclass[runningheads]{llncs}
\usepackage{graphicx}
\usepackage{subcaption}
\usepackage{hyperref, xcolor}
\usepackage{siunitx}
\usepackage{amssymb}
\begin{document}

{\let\thefootnote\relax\footnotetext{Copyright \textcopyright\ 2021 for this paper by its authors. Use permitted under Creative Commons License Attribution 4.0 International (CC BY 4.0).}}

\title{On the Role of Images for Analyzing Claims in Social Media}

\titlerunning{Multimodal Check-worthy Claim Detection}

\author{Gullal S. Cheema\inst{1}\orcidID{0000-0003-4354-9629} \and
Sherzod Hakimov\inst{1}\orcidID{0000-0002-7421-6213} \and
Eric M{\"u}ller-Budack\inst{1}\orcidID{0000-0002-6802-1241} \and Ralph Ewerth\inst{1,2}\orcidID{0000-0003-0918-6297}}

\authorrunning{Cheema et al.}

\institute{TIB -- Leibniz Information Centre for Science and Technology, Hannover, Germany
\\
\and
L3S Research Center, Leibniz University Hannover, Germany\\
\email{\{gullal.cheema,sherzod.hakimov,eric.mueller,ralph.ewerth\}@tib.eu}}
\maketitle              
\begin{abstract}
Fake news is a severe problem in social media. In this paper, we present an empirical study on visual, textual, and multimodal models for the tasks of claim, claim check-worthiness, and conspiracy detection, all of which are related to fake news detection. Recent work suggests that images are more influential than text and often appear alongside fake text. To this end, several multimodal models have been proposed in recent years that use images along with text to detect fake news on social media sites like \emph{Twitter}. However, the role of images is not well understood for claim detection, specifically using transformer-based textual and multimodal models. We investigate state-of-the-art models for images, text (Transformer-based), and multimodal information for four different datasets across two languages to understand the role of images in the task of claim and conspiracy detection.

\end{abstract}

\section{Introduction}

Social media platforms have become an integral part of our everyday lives, where we use them to connect with people and consume news, entertainment, and buy or sell products. In the last decade, social media has seen exponential growth, with more than a couple of billion users and the increasing presence of prominent people like politicians and celebrities (also called Influencers), organizations, and political parties. On the one hand, this allows influential people or organizations to reach millions of users directly, but it also allows for fake and unverified information to rise and spread faster~\cite{vosoughi2018spread} due to the nature of social media. To deal with misinformation and false claims on online platforms, several independent fact-checking projects like \textit{Snopes, Alt News, Our.News} have been launched that manually fact-check news and publish their outcomes for public use. Although more such initiatives are coming up worldwide, they cannot keep up with the rate of news or information production on online platforms. Therefore, fake news detection has gathered much interest in computer science for developing automated methods to speed and scale up to handle the continuous fast streaming social media data.


As social media is inherently multimodal in nature, fact-checking initiatives and computation methods consider not only text but also image content~\cite{giachanou2019leveraging,khattar2019mvae,singhal2019spotfake,wang2018eann} as it can be easily fabricated and manipulated due to the availability of free image and video editing tools. In this paper, we investigate the role of images in the context of claim and conspiracy detection. Claim detection is one of the first vital steps to identify fake news where the purpose is to flag a statement if it contains check-worthy facts and information, while the claim may be true or false. Whereas in conspiracy detection, a statement that includes a conspiracy theory is fake news and consists of manipulated facts. Although fake news on social media has been explored recently from a multimodal perspective, images have hardly been considered for claim detection except in recent work by Zlatkova \textit{et al.}~\cite{zlatkova2019fact}. Here, meta-information of images is treated as features, and reverse image search is performed to compare the claim text. However, the image's semantic information is not considered, and the authors highlight that images are more influential than text and appear alongside fake text or unverified news.

Since we are interested in the impact of using images in a multimodal framework, to keep our models simple, we focus on extracting only semantic or contextual features from text and do not consider its structure or syntactic information. To this end, we mainly consider deep transformer Bidirectional Encoder Representations from Transformers (BERT) to extract contextual embeddings and use them along with image embeddings. Taking inspiration from recent work by Cao \textit{et al.}~\cite{cao2020exploring}, we extract image sentiment features that are widely applied for image credibility or fake news detection in addition to object and scene information for the semantic overlap with textual information.

\begin{figure}[t]
	\centering
	\centering
	\includegraphics[width=1.0\textwidth]{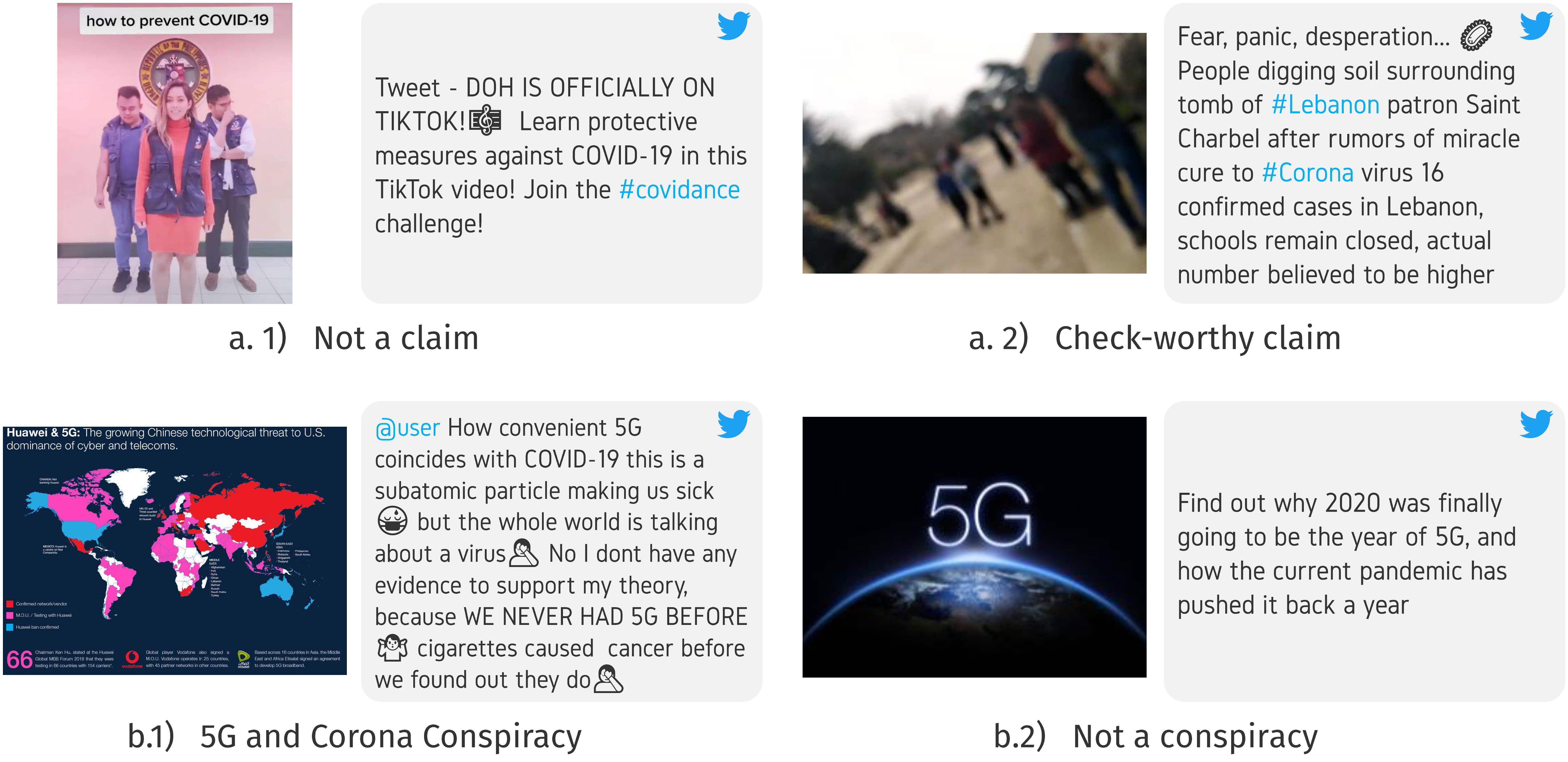}
	\label{fig:clef_en}
	\caption{Examples from CLEF-English~\cite{clef-checkthat-en:2020} (a) check-worthy claims dataset and MediaEval~\cite{pogorelov2020fakenews} (b) conspiracy detection dataset}
	\label{fig:intro_ex}
\end{figure}

To carry out this study\footnote{Code: \url{https://github.com/cleopatra-itn/image_text_claim_detection}}, we experiment with four \emph{Twitter} datasets\footnote{Dataset: \url{https://zenodo.org/record/4592249}} on binary classification tasks, two of which are from the recent \emph{CLEF-CheckThat! 2020}~\cite{barron2020overview}, one in English~\cite{clef-checkthat-en:2020} and the other one in Arabic~\cite{clef-checkthat-ar:2020}. The third one is an English dataset from \emph{MediaEval 2020}~\cite{pogorelov2020fakenews} on conspiracy detection, and the last one is a recent claim detection dataset (English) from Gupta \textit{et al.}~\cite{gupta2021lesa} on \emph{COVID-19} tweets. Four examples for claim and conspiracy detection are shown in Figure~\ref{fig:intro_ex}. To train our unimodal and multimodal models, we use Support Vector Machines (SVM)~\cite{suykens1999least} and Principal Component Analysis (PCA)~\cite{wold1987principal} for dimensionality reduction due to the small datasets and large size of combined features. We also fine-tune BERT models on the text input to see the extent of the unimodal model's performance on limited-sized datasets and use different pre-trained \emph{BERT} models to see the effect of domain gap. Furthermore, we investigate the recently proposed transformer-based \emph{ViLBERT}~\cite{lu2019vilbert} (Vision-and-Language BERT) model that learns semantic features via co-attention on image and textual inputs. Just like \emph{BERT} models, we perform fixed embedding and fine-tuning experiments using \emph{ViLBERT} to see if a large transformer-based multimodal model can learn meaningful representation and perform better on small-sized datasets.

The remainder of the paper is organized as follows. Section~\ref{sec:related_work} briefly discusses related work on fake news detection and the sub-problems of claim and conspiracy detection. Section~\ref{sec:method} presents details of image, text, and multimodal features as well as the fine-tuned and applied models. Section~\ref{sec:experiments} describes the experimental setup, results and summarizes our findings. Section~\ref{sec:conclusion} concludes the paper with future research directions.

\section{Related Work}\label{sec:related_work}

There is a wide body of work on fake news detection that goes well beyond this paper's scope. Therefore, we restrict this section to multimodal fake news, claim detection, and conspiracy detection. 

\subsection{Unimodal Approaches}
The earliest claim detection works go back a decade. Rosenthal \textit{et al.}~\cite{rosenthal2012detecting} in their pioneering work extracted claims from \emph{Wikipedia} discussion forums. They classified them via logistic regression using the sentiment, syntactic and lexical features like POS (Part-of-Speech) tags and n-grams, and other statistical features over text. Since then, researchers have proposed context dependent~\cite{levy2014context}, context independent~\cite{lippi2015context}, and cross-domain~\cite{daxenberger2017essence} and in-domain approaches for claim detection. Recently, the transformer-based models~\cite{chakrabarty2019imho} have replaced structure-based claim detection approaches due to their success in several downstream natural language processing (NLP) tasks.

For claim detection on social media in particular, recently \emph{CLEF-CheckThat! 2020}~\cite{barron2020overview} hosted a challenge to detect check-worthy claims in \emph{COVID-19} related English tweets and several other topics in Arabic. The challenge attracted several models with top submissions~\cite{cheema2020checksquare,clef-checkthat-Nikolov:2020,clef-checkthat-williams:2020} all using some version of transformer-based models like \emph{BERT}~\cite{devlin2018bert} and \emph{RoBERTa}~\cite{liu2019roberta} along with tweet meta-data and lexical features. Outside of CLEF challenges, some works~\cite{dogan2015detecting,majithia2019claimportal} have also conducted a detailed study on detecting check-worthy tweets in U.S. politics and proposed real-time systems to monitor and filter them. Taking inspiration from~\cite{daxenberger2017essence}, Gupta \textit{et al.}~\cite{gupta2021lesa} address the limitations of current methods in cross-domain claim detection by proposing a generalized claim detection model called \emph{LESA} (Linguistic Encapsulation and Semantic Amalgamation). Their model combines contextual transformer features with learnable POS and dependency relation embeddings via transformers to achieve impressive results on several datasets. For conspiracy detection, \emph{MediaEval 2020}~\cite{pogorelov2020fakenews} saw interesting methods to automatically detect 5G and Coronavirus conspiracy in tweets. Top submissions used BERT~\cite{cheema2021tib,tuan2020mediaeval} pre-trained on \emph{COVID} Twitter data, tweet meta-data, graph network data and RoBERTa models~\cite{claveau2020mediaeval} along with Graph Convolutional Neural (GCN) networks.

\vspace{-0.2cm}
\subsection{Multimodal Approaches}
For multimodal fake news in general, several benchmark datasets have been proposed in the last few years, generating interest in developing multimodal visual and textual models. In one of the relatively early works, Jin \textit{et al.}~\cite{jin2017multimodal} explored rumor detection on Twitter using text, social context (emoticons, URLs, hashtags), and the image by learning a joint representation with attention from LSTM outputs over image features. The authors observed the benefit of using the image and social context in addition to text by improving the detection of fake news in Twitter and Weibo datasets. Later, Wang \textit{et al.}~\cite{wang2018eann}, proposed an improved model that learns a multi-task model to detect fake news as one task and event discriminator as another task to learn event invariant representations. Since then, improvements have been proposed via using multimodal variational autoencoders~\cite{khattar2019mvae}, transfer learning~\cite{giachanou2020multimodal,singhal2019spotfake} with transformer-based text and deep visual CNN models. Recently, Nakamura~\cite{nakamura2020fakeddit} \textit{et al.} proposed a fake news dataset \textit{r/Fakeddit} mined from Reddit with over 1 million samples, which includes text, images, meta-data, and comments data. The data is labeled through distant supervision into 2-way, 3-way, and 6-way classification categories. In addition to our different tasks, another difference with the approaches mentioned above is that the size of the datasets is moderate (several thousand) to large (millions) in comparison to a few hundred or a couple of thousand samples in our four datasets for \textit{claim} and \textit{conspiracy detection}.

\section{Methodology}\label{sec:method}
\begin{figure}[t]
	\centering
	\includegraphics[width=\textwidth]{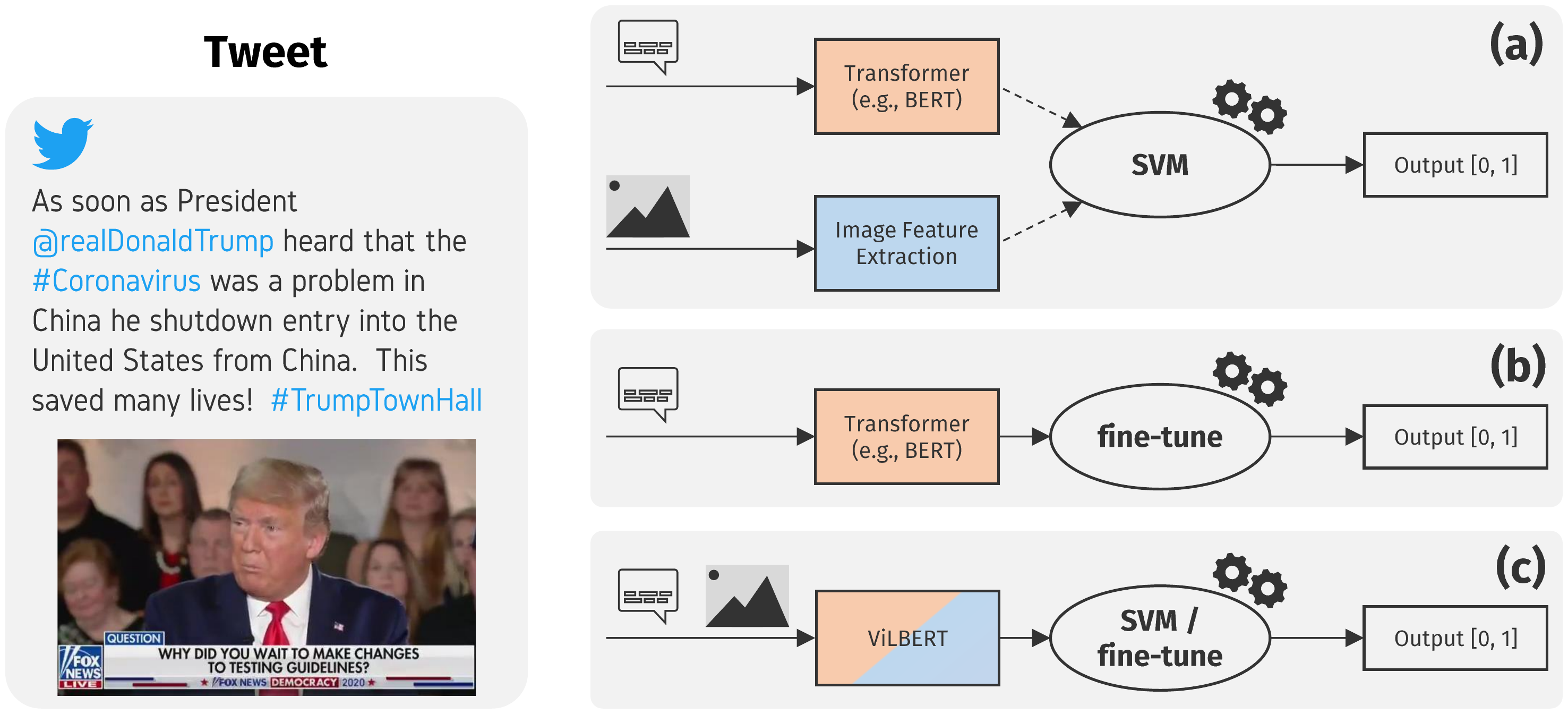}
	\caption{Workflow of the proposed solutions for claim and conspiracy detection in tweets using multimodal information from text~(orange) and image~(blue). Three approaches are investigated: (a)~Training of an SVM based on the combined features extracted from pre-trained models for visual and/or textual information extraction. (b)~Fine-tuning a transformer network (\emph{BERT}) solely using textual information. (c)~Fine-tuning \textit{ViLBERT}~\cite{lu2019vilbert} or training an SVM based on its multimodal embeddings extracted from text and image.}
	\label{fig:workflow}
\end{figure}

In this section, we provide details of different image (Section~\ref{subsec:image_model}), textual (Section~\ref{subsec:textual_model}), and multimodal (Section~\ref{subsec:multimodal_model}) models and their feature encoding process and how classification models (Section~\ref{subsec:classification}) are built. An overview of classification models are presented in Figure~\ref{fig:workflow}.

\subsection{Image Models ($I$)}\label{subsec:image_model}
The purpose of image models is to encode the presence of different objects, scene, place or background, and affective image content. When learning a multimodal model or a classifier, specific overlapping patterns between image and text can act as discriminatory features for claim detection. 
\\
\noindent\textbf{Object Features ($I_o$)}
In order to encode objects and the overall image content, we extract features from a pre-trained \emph{ResNet}~\cite{he2016deep} model trained on \textit{ImageNet}~\cite{russakovsky2015imagenet} dataset. The pre-trained model has been shown to boost performance over low-level features in several computer vision tasks. We use widely recognized \emph{ResNet}-152 and its last convolution layer to extract features instead of the object categories (final layer). The final convolutional layer outputs 2048 feature maps each of size $7 \times 7$, which is then pooled with a global average to get a 2048-dimensional vector.
\\
\noindent\textbf{Place and Scene Features ($I_p$)}
In order to encode the scene information in an image, we extract features from a pre-trained \emph{ResNet}~\cite{he2016deep} model trained on \textit{Places365}~\cite{zhou2017places} dataset. In this case, we use \emph{ResNet}-101 and follow the same encoding process as described for object features.
\\
\noindent\textbf{Hybrid Object and Scene Features ($I_h$)}
We also experiment with a hybrid model trained on both \textit{ImageNet} and \textit{Places365} datasets that encodes object and scene information in a single model. To extract these features, we again use a \emph{ResNet}-101 model and follow the same encoding process.
\\
\noindent\textbf{Image Sentiment ($I_s$)}
To encode the image sentiment, we use a pre-trained model~\cite{Vadicamo_2017_ICCVW} that is trained on three million images using weak supervision of sentiment label from the tweet text. Although the image labels are noisy, the model has shown superior performance on unseen Twitter testing datasets. We use their best CNN model based on \emph{VGG-19}~\cite{simonyan2014very}. The image sentiment embeddings ($I_{se}$) are extracted from the last layer in the model, which are 4096-dimensional vectors. Additionally, we extract the image sentiment predictions ($I_{sp}$) from the classification layer that outputs a three-dimensional vector corresponding to the probabilities of three sentiment classes (Negative, Neutral and Positive).

\subsection{Textual Models ($T$)}\label{subsec:textual_model}
Since context and semantics of the sentence is shown~\cite{barron2020overview,chakrabarty2019imho} to be important for claim detection, we use transformer-based \emph{BERT}-Base~\cite{devlin2018bert} ($T_{BB}$), to extract contextual word embeddings and employ different pooling strategies to get a single embedding for the tweet. As different layers of \emph{BERT} capture different kinds of information, we experiment with four combinations, i.e., 1) concatenate the last four hidden layers, 2) sum of the last four hidden layers, 3) the last hidden layer,  and 4) the second last hidden layer. We finally take an average over the word embeddings to obtain a single vector.

To reduce the domain gap for our Twitter datasets in English, we experiment with two \emph{BERT} models. The first variant is called \emph{BERTweet}~\cite{bertweet} ($T_{BT}$) a BERT-base model that is further pre-trained on 850 million English tweets, and the second one called \emph{COVID-Twitter-BERT}~\cite{muller2020covid} ($T_{CT}$), a BERT-large model trained on 97 million English tweets on the topic of \textit{COVID-19}. For Arabic tweets, we experiment with the \emph{AraBERT}~\cite{antoun2020arabert} ($T_{AB}$) that is trained on Arabic news corpus called \emph{OSIAN}~\cite{zeroual-etal-2019-osian} and 1.5 Billion words Arabic corpus~\cite{ElKhair201615BW}. We also perform two experiments, one with raw tweets and the other with pre-processing tweets as part of the \emph{AraBERT's} language-specific text processing method. 

For English text, with vanilla BERT-base model, we pre-process the text by following the steps mentioned in Cheema \textit{et. al.}~\cite{cheema2020checksquare} using the publicly available text processing tool Ekphrasis~\cite{baziotis-pelekis-doulkeridis:2017:SemEval2}. We also show the performance of vanilla \emph{BERT}-base on raw tweets ($T_{BB}^{Raw}$) to reflect its sensitivity towards text pre-processing ($T_{BB}^{Clean}$). For both \emph{BERTweet} and \emph{COVID-Twitter-BERT}, we follow their pre-processing steps, which normalize text, and additionally replaces \textit{user mentions, emails, URLs} with special keywords.

\subsection{Multimodal Models ($M$)}\label{subsec:multimodal_model}

\subsubsection{ViLBERT (Vision-and-Language BERT)}
We use \emph{ViLBERT}~\cite{lu2019vilbert}, one of the recent multimodal transformer architectures that process image and text inputs through two separate transformer-based streams and combines them through transformer layers with the co-attention. It eventually outputs co-attended image and text features that can be combined (added, multiplied or concatenated) to learn a classifier for vision and language tasks. The authors proposed to use visual grounding as a self-supervised pre-training task on a large conceptual captions dataset~\cite{sharma2018conceptual}. They used the model for various downstream tasks involving vision and language, such as visual question answering, visual commonsense reasoning, and caption-based image retrieval. 

For the image branch, \emph{ViLBERT} uses state-of-the-art object detection model \emph{Mask R-CNN}~\cite{he2017mask} and extracts top 100 region proposals (boxes) and their corresponding features. These features are used in a sequence through a 5-layer image transformer, which outputs the image region embeddings. For the text branch, it uses BERT-base model to get the contextual word embeddings. A 6-layer transformer block with the co-attention follows the individual streams that outputs the co-attended image and text embeddings. 

\subsubsection{Feature Extraction}
In our fixed embedding experiments with a SVM, we experiment with the output of pooling and last layers of image and text branches. With pooling layers, we directly concatenate ($M_{pool}^{CAT}$) the image and text outputs. With last layer outputs we average the image region embeddings and word embeddings to get one single embedding per modality and then concatenate them ($M_{avg}^{CAT}$). From pooling layers, each modality's embedding size is a 1024-dimensional vector, and the last layer average of embeddings gives 1024 and 768-dimensional vectors for image and text, respectively. For fine-tuning, we follow \emph{ViLBERT}'s downstream task approach, where the pooling layer outputs are either added ($M_{pool}^{ADD}$) or multiplied ($M_{pool}^{MUL}$) and passed to a classifier. For Arabic text, we use Google Translate to convert the text into English because all \emph{ViLBERT} models are trained on English text.

\emph{ViLBERT} is fine-tuned on several downstream tasks which can be relevant for encapsulating image-text relationship for our claim detection problem. Therefore, we experiment with four different pre-trained models, namely, conceptual captions , image retrieval (\emph{Image-Ret}), grounding referring expressions (localize an image region given a natural language reference) (\emph{RefCOCO}), and a multi-task model~\cite{Lu_2020_CVPR} that is trained on 12 different tasks. 

\subsection{Classification of Tweets}\label{subsec:classification}
For our fixed embedding experiments, we train SVM models with each type of image and text embeddings for binary classification of tweets as shown in Figure~\ref{fig:workflow}~(a). For fine-tuning textual models (Figure~\ref{fig:workflow}~(b)), given that we have relatively small-sized datasets, we only experiment with fine-tuning the last two and four layers of transformer models for each dataset. We concatenate the image and text features for multimodal fixed embedding experiments and train an SVM model over them for classification.

In the case of \emph{ViLBERT} (Figure~\ref{fig:workflow}~(c)), we again train SVM over the extracted pooled image and text outputs for classification. For fine-tuning, we fix the individual transformer branches and experiment with fine-tuning the last two and four co-attention layers to activate the interaction between modalities. It enables us to see the effect of only the attention mechanism that can show the benefit of an image and text in claim detection. We use a simple classifier on top of \emph{ViLBERT} outputs as recommended by the authors of \emph{ViLBERT}, which includes a linear layer for down projecting outputs to 128 dimensions, followed by \emph{ReLU} (Rectified Linear Unit) non-linear activation function, a normalization layer and finally a binary classification layer. Dropout is used to avoid over-fitting, and the fine-tuning is performed by minimizing the cross-entropy loss.






\section{Experiments and Results}\label{sec:experiments}

In this section, we describe all the datasets and their statistics, training details and hyper-parameters, model details, experimental results, and discuss them as obtained by different models mentioned in Section~\ref{sec:method}. 

\subsection{Datasets}
We selected the following four publicly available \emph{Twitter} datasets with high-quality annotations (which excludes~\cite{nakamura2020fakeddit}, besides its focus on fake news), three of which are on claim detection and one on conspiracy detection.
The number of tweets in the original datasets is four to fifteen times more as they were mined for text-based fake news detection. We only selected tweets that have an image.

\noindent\textbf{CLEF-En~\cite{clef-checkthat-en:2020}} - Released as a part of \emph{CLEF-CheckThat! 2020} challenge, the purpose is to identify \emph{COVID-19} related tweets that are check-worthy claims vs not check-worthy claims. Only 281 English tweets in the dataset include images, whereas the original dataset included 964 tweets.

\noindent\textbf{CLEF-Ar~\cite{clef-checkthat-ar:2020}} - Released in the same challenge, the dataset consists of 15 topics related to middle east including \emph{COVID-19} and the purpose is to identify check-worthy claims. It consists of 2571 Arabic tweets and corresponding images.

\noindent\textbf{MediaEval~\cite{pogorelov2020fakenews}} - Released in \emph{MediaEval 2020} workshop~\cite{pogorelov2020fakenews} challenge on identifying 5G and Coronavirus conspiracy tweets. The original dataset has three classes, 5G and Corona conspiracy, other conspiracies, and no conspiracy. To make the problem consistent with other datasets in this paper, we combine conspiracy classes (Corona and others) and treat it as a binary classification problem. It consists of \num{1724} tweets and images.

\noindent\textbf{LESA~\cite{gupta2021lesa}} - This is a recently proposed dataset of \emph{COVID-19} related tweets on the problem of claim detection. Here, the problem is identifying whether a tweet is a claim or not, and not the claim check-worthiness as in \emph{CLEF-En}.
The original dataset consists of \num{10000} tweets in English, out of which only \num{1395} consists of images.

We applied 5-fold cross-validation to overcome the issue of low number of samples in each dataset. We used the ratio of around 72:10:18 for training, validation, and testing in each data split. Next, we report the experimental results for different model configurations. The reported results are averaged across five splits of each dataset. We report accuracy and weighted-F1 measure to account for label imbalance in all the datasets.

\subsection{Setup and Hyper-parameters}\label{subsec:exp_features}
\noindent \textbf{SVM hyper-parameters}: we perform grid search over PCA energy (\%) conservation, regularization parameter \textit{C} and RBF kernel's \textit{gamma}. The parameter range for \textit{PCA} varies from 100\% (original features) to 95\% with decrements of 1. The parameter range for \textit{C} and \textit{gamma} vary between \num{-1} to \num{1} on a log-scale with 15 steps. For experiments only on the \emph{CLEF-En} dataset, we use the range between \num{-2} to \num{0} for \textit{C} and \textit{gamma}, as the number of samples are very low and needs aggressive regularization. We normalize the final embedding so that $l2$ norm of the vector is \num{1}.

\noindent\textbf{Fine-tuning BERT and VilBERT}: we use a batch size of 4 for \emph{CLEF-En} and 16 for the other datasets. We train all the models for 6 epochs with a starting learning rate of $5e-5$ and a linear decay.  A dropout with ratio \num{0.2} is applied after the first linear layer in the classifier for regularization during fine-tuning.

\bgroup
\setlength{\tabcolsep}{2pt} 
\renewcommand{\arraystretch}{1.2} 
\begin{table}[h]
\centering
\caption{The classification results on all datasets using the textual and visual features (see Sections~\ref{subsec:image_model} and \ref{subsec:textual_model}). Models marked with$^{\dagger}$ are fine-tuning results and the rest are SVM-based. The best result for each group (bold), and the best result for each dataset (bold and underlined) are highlighted. }

\begin{tabular}{|l|c|c|c|c|c|c|c|c|}
\hline
\multicolumn{1}{|c|}{\textbf{Model}} & \multicolumn{2}{c|}{\textbf{CLEF-En}~\cite{clef-checkthat-en:2020}} & \multicolumn{2}{c|}{\textbf{CLEF-Ar}~\cite{clef-checkthat-ar:2020}} & \multicolumn{2}{c|}{\textbf{LESA}~\cite{gupta2021lesa}} & \multicolumn{2}{c|}{\textbf{MediaEval}~\cite{pogorelov2020fakenews}} \\ \hline
& \textbf{ACC}      & \textbf{F1}       & \textbf{ACC}      & \textbf{F1}       & \textbf{ACC}     & \textbf{F1}     & \textbf{ACC}       & \textbf{F1}        \\ \hline
\textbf{$I_o$}                        & \textbf{0.6748}   & \textbf{0.6180}   & 0.6991            & 0.6770            & \textbf{0.8223}  & \textbf{0.7775} & 0.7189             & \textbf{0.6968}             \\ \hline
\textbf{$I_p$}                        & 0.6033            & 0.5966            & 0.6961            & 0.6558            & 0.8159           & 0.7687          & 0.7215             & 0.6355             \\ \hline
\textbf{$I_h$}                        & 0.6551            & 0.6108            & \textbf{0.7052}   & \textbf{0.6776}   & 0.8223           & 0.7744          & \textbf{0.7260}    & 0.6581    \\ \hline
\textbf{$I_{se}$}                       & 0.6384            & 0.6223            & 0.7073            & 0.6563            & 0.8143           & 0.7516          & 0.708              & 0.6822             \\ \hline
\hline
\textbf{$T_{BB}^{Clean}$}               & 0.7501            & 0.7514            & -                 & -                 & \textbf{0.8279}  & \textbf{0.8015} & 0.8130             & 0.8026             \\ \hline
\textbf{$T_{BB}^{Raw}$}                 & 0.7459            & 0.7346            & -                 & -                 & 0.8119           & 0.7873          & 0.8298             & 0.8261             \\ \hline
\textbf{$T_{BT}$}                     & \underline{\textbf{0.7656}}   & \underline{\textbf{0.7661}}  & -                 & -                 & 0.8255           & 0.8023          & 0.8272             & 0.8232             \\ \hline
\textbf{$T_{CT}$}                     & 0.7178            & 0.7123            & -                 & -                 & 0.8175           & 0.8045          & \textbf{0.8479}    & \textbf{0.8479}    \\ \hline
\textbf{$T_{AB}$}                   & -                 & -                 & \textbf{0.8362}   & \textbf{0.8307}   & -                & -               & -                  & -                  \\ \hline
\hline
\textbf{$T{_{BB}^{Clean}}^{\dagger}$}              & 0.6942            & 0.6804            & -                 & -                 & 0.8319           & 0.809           & 0.8046             & 0.7952             \\ \hline
\textbf{$T_{BT}^{\dagger}$}                    & \textbf{0.7420}   & \textbf{0.7363}   & -                 & -                 & \underline{\textbf{0.8486$^2$}}  & \underline{\textbf{0.8303$^2$}} & 0.8407$^2$             & 0.8342$^2$             \\ \hline
\textbf{$T_{CT}^{\dagger}$}                     & 0.7146            & 0.6784            & -                 & -                 & 0.8303           & 0.8075          & \underline{\textbf{0.8627}}    & \underline{\textbf{0.8604}}    \\ \hline
\textbf{$T_{AB}^{\dagger}$}                  & -                 & -                 & \underline{\textbf{0.8431}}   & \underline{\textbf{0.8432}}   & -                & -               & -                  & -                  \\ \hline
\end{tabular}
\label{tab:uni_res}
\end{table}
\egroup

\subsection{Results}
Table~\ref{tab:uni_res} and Table~\ref{tab:multi_res} show the unimodal and multimodal models' performance for all the four datasets based on type of features and feature combinations respectively.

\emph{\textbf{Unimodal Results}} - In Table~\ref{tab:uni_res}, it can be seen that all the visual features perform poorly in comparison to textual features. This is expected as visual information on its own cannot indicate whether a social media post makes a claim unless it has text or it's a video. Among the four types of visual models, \emph{Object} ($I_o$) and \emph{Hybrid} ($I_h$) features are slightly better, probably because the place or scene information (lowest F1 for all datasets) on its own is not a useful indicator in images for claim detection. With textual features, \emph{BERT} models that are further pre-trained on tweets ($T_{BT}, T_{BT}^{\dagger}$) and \emph{COVID}-related data ($T_{CT}, T_{CT}^{\dagger}$) perform better in comparison to vanilla \emph{BERT} ($T_{BB}^{Clean}, T{_{BB}^{Clean}}^{\dagger}$) in at-least three datasets. It suggests that the tweets' structure and the domain gap are better captured and reduced respectively in Twitter corpus pre-trained models. Further, normalizing ($T_{BB}^{Clean}$) the tweet text delivers better performance than using the raw text ($T_{BB}^{Raw}$). In SVM training, we observed the sum of the last four layers of BERT to compute the embeddings performs better than the other pooling combinations. It indicates that downstream tasks can benefit from the diverse information in different layers of BERT. Similarly, fine-tuning the last four layers instead of two (marked with$^2$) gives better performance across all the datasets with \emph{BERT-base} ($T_{BB}^{Clean}$$^{\dagger}$), \emph{COVID-Twitter-BERT} ($T_{CT}$$^{\dagger}$) and \emph{AraBERT} ($T_{AB}$$^{\dagger}$).



\emph{\textbf{Multimodal Results}} - In Table~\ref{tab:multi_res}, we can see the effect of combining visual features with textual features by using a simple concatenation in SVM and also with multimodal co-attention transformer \emph{ViLBERT}. Although we do not see any benefit of using the image sentiment embeddings ($I_{se}$) in unimodal models, here instead, we use the image sentiment predictions ($I_{sp}$) that perform better or equivalent in comparison to other visual features. For instance, in case of \emph{CLEF-Ar}, sentiment predictions $I_{sp}$ with \emph{AraBERT} ($T_{AB}$$^{\dagger}$) gives the best fixed embedding performance. Similarly, combining hybrid features ($I_h$) with \emph{BERT-base} ($T_{BB}^{Clean}$$^{\dagger}$) and object features with \emph{COVID-Twitter-BERT} ($T_{CT}$$^{\dagger}$) in case of \emph{LESA} and \emph{MediaEval} improves the metrics by 1\% over textual SVM models. 

With \emph{ViLBERT}, it is interesting to see that with fixed visual and textual branches, it can capture some information from image and text with co-attention to boost performance in case of \emph{LESA} and \emph{MediaEval}. It is worth mentioning that the best unimodal textual models for English and Arabic are pre-trained models further trained on Twitter and language-specific data corpus. In the case of \emph{ViLBERT}, there is a wider domain gap, and for Arabic, the translation process loses quite a bit of information that results in a drop in performance. Different pooling operations applied for pre-trained \emph{ViLBERT} models show more difference in fixed-embedding SVM experiments where the average pooling ($M_{avg}^{CAT}$) yields a considerable performance, which we also observed in unimodal SVM experiments. We observed that pre-training tasks (best two reported in Table~\ref{tab:multi_res}) also matter, where image retrieval (Image-Ret) and language reference grounding (RefCOCO) features perform much better for all the datasets. It is explainable since both tasks require capturing complex relationships and linking text to specific image regions in the image, enabling them to perform better for our tasks.

\bgroup
\setlength{\tabcolsep}{2pt} 
\renewcommand{\arraystretch}{1.2} 
\begin{table}[t]
\centering
\caption{Multimodal classification results on all datasets based on combination of textual, visual and multimodal features (see Sections~\ref{subsec:image_model}, \ref{subsec:textual_model} and \ref{subsec:multimodal_model}). Models marked with$^{\dagger}$ are fine-tuning results on VilBERT and the rest are SVM-based. \emph{ReCOCO} and \emph{Image-Ret} refers to pre-trained \emph{ViLBERT} models. Layers refers to the number of fine-tuned co-attention layers in VilBERT.}
\begin{tabular}{|l|c|c|c|c|c|c|c|c|}
\hline
\multicolumn{1}{|c|}{\textbf{Model}} & \multicolumn{2}{c|}{\textbf{CLEF-En}~\cite{clef-checkthat-en:2020}} & \multicolumn{2}{c|}{\textbf{CLEF-Ar}~\cite{clef-checkthat-ar:2020}} & \multicolumn{2}{c|}{\textbf{LESA}~\cite{gupta2021lesa}} & \multicolumn{2}{c|}{\textbf{MediaEval}~\cite{pogorelov2020fakenews}} \\ \hline
                                     & \textbf{ACC}                 & \textbf{F1} & \textbf{ACC}       & \textbf{F1}      & \textbf{ACC}     & \textbf{F1}     & \textbf{ACC}        & \textbf{F1}       \\ \hline
\hline
\textbf{Best Unimodal}                                        & 0.7656             & 0.7661            & 0.8431             & 0.8432             & 0.8486             & 0.8303            & 0.8627    & 0.8604  \\ \hline \hline
\multicolumn{1}{|c|}{\textbf{$T$ -\textgreater{}}} & \multicolumn{2}{c|}{\textbf{$T_{BB}^{Clean}$}} & \multicolumn{2}{c|}{\textbf{$T_{AB}$}~\cite{antoun2020arabert}}   & \multicolumn{2}{c|}{\textbf{$T_{BB}^{Clean}$}} & \multicolumn{2}{c|}{\textbf{$T_{CT}$}~\cite{muller2020covid}}        \\ \hline
\textbf{$I_o$ + $T$}                                        & 0.7219             & 0.7053            & 0.8054             & 0.8053             & 0.8311             & 0.7953            & \underline{\textbf{0.8594}}    & \underline{\textbf{0.8566}}    \\ \hline
\textbf{$I_p$ + $T$}                                        & 0.7336             & 0.7296            & 0.8184             & 0.8168             & 0.8223             & \textbf{0.7955}            & 0.8472             & 0.8460             \\ \hline
\textbf{$I_h$ + $T$}                                        & 0.7259             & 0.7003            & 0.8085             & 0.8060             & \textbf{0.8335}    & 0.7907   & 0.8549             & 0.8527             \\ \hline
\textbf{$I_{sp}$ + $T$}                                       & \underline{\textbf{0.7557}}    & \underline{\textbf{0.7575}}   & \underline{\textbf{0.8370}}    & \underline{\textbf{0.8319}}    & 0.8271             & 0.8009            & 0.8485             & 0.8483             \\ \hline
\hline
\multicolumn{1}{|c|}{\textbf{Pre-trained -\textgreater{}}} & \multicolumn{2}{c|}{\textbf{RefCOCO}}  & \multicolumn{2}{c|}{\textbf{Image-Ret}} & \multicolumn{2}{c|}{\textbf{RefCOCO}}  & \multicolumn{2}{c|}{\textbf{Image-Ret}} \\ \hline
\textbf{$M_{pool}^{CAT}$}                                       & 0.6980             & 0.6941            & 0.7125             & 0.6990             & 0.8175             & 0.7842            & 0.7357             & 0.7142             \\ \hline
\textbf{$M_{avg}^{CAT}$}                                       & \textbf{0.7062}    & \textbf{0.7022}   & \textbf{0.7454}    & \textbf{0.7245}    & \textbf{0.8175}    & \textbf{0.7910}   & \textbf{0.7892}    & \textbf{0.7832}    \\ \hline
\hline
\multicolumn{1}{|c|}{\textbf{Layers -\textgreater{}}}       & \multicolumn{2}{c|}{\textbf{2}}        & \multicolumn{2}{c|}{\textbf{4}}         & \multicolumn{2}{c|}{\textbf{4}}        & \multicolumn{2}{c|}{\textbf{4}}         \\ \hline
\textbf{$M{_{pool}^{ADD}}^{\dagger}$}                                     & \textbf{0.7339}    & \textbf{0.7322}   & 0.7449             & 0.7214             & \underline{\textbf{0.8446}}    & \underline{\textbf{0.8196}}   & 0.7989             & 0.7820             \\ \hline
\textbf{$M{_{pool}^{MUL}}^{\dagger}$}                                      & 0.7336             & 0.7341            & \textbf{0.7449}    & \textbf{0.7389}    & 0.8446             & 0.8191            & 0.7937             & 0.7772             \\ \hline
\textbf{$M{_{avg}^{CAT}}^{\dagger}$}                                       & 0.7182             & 0.7121            & 0.7466             & 0.7415             & 0.8319             & 0.8063            & \textbf{0.8014}    & \textbf{0.7900}    \\ \hline
\end{tabular}
\label{tab:multi_res}
\end{table}
\egroup

\subsection{Discussion of Results}
We can summarize the findings of our experiments as follows: \textbf{1) Domain-specific languages models should be preferred for downstream tasks such as claim detection or fake news}, where underlying meaning and context of certain words (like COVID) is essential, \textbf{2) Multimodality certainly helps as seen with multimodal transformer models}, where activating interaction through co-attention layers between fixed unimodal embeddings improves the performance in two datasets, \textbf{3) To further understand underlying multimodal dynamics it might be better to explicitly model multimodal relationships}, for instance, importance of image or correlation between image-text in addition to claim detection, \textbf{4) Certain pre-training tasks in \emph{ViLBERT} are better suited for downstream tasks} and need further introspection on larger datasets, and lastly, \textbf{5) Visual models need to be better adapted to social media images}, for instance, the models used here are not sufficient for diagrams or images with large text, which constitute around 30-40\% of \emph{LESA} and \emph{MediaEval} datasets.

\section{Conclusion}\label{sec:conclusion}

In this paper, we have investigated the role of images and tweet text for two problems related to fake news, claim, and conspiracy detection. For this purpose, we combined several state-of-the-art CNN features for images with BERT features for text. We observed the performance improvement over unimodal models in two out of four \emph{Twitter} datasets over two languages. We also experimented with the recently proposed multimodal co-attention transformer \emph{ViLBERT} and observed a promising performance using both image and text even with relatively small-sized datasets. In future work, we will look into other ways to include external knowledge in domain-independent claim detection models without relying on different domain-specific language models. Second, we plan to investigate multimodal transformers in more detail and analyze if the performance does scale with more data in similar tasks. Finally, to address the limitation of visual models, we will consider models that can deal with text and graphs in images and extract suitable features.


\section*{Acknowledgements}
This work was funded by European Union’s Horizon 2020 research and innovation programme under the Marie Skłodowska-Curie grant agreement no 812997.

%
%

\bibliographystyle{splncs04}
\bibliography{main}
\end{document}